\RequirePackage{fix-cm}
\documentclass{svjour3}                     
\smartqed  
\usepackage{graphicx}

\usepackage{amsmath,amssymb}
\usepackage{bm}
\usepackage[hang,small,bf]{caption}
\usepackage{enumitem}
\usepackage{float}
\usepackage[margin=30truemm]{geometry}
\usepackage[utf8]{inputenc}
\usepackage{mathtools}
\usepackage[subrefformat=parens]{subcaption}
\usepackage{url}
\usepackage{color}
\usepackage{comment}
\usepackage[table]{xcolor}

\makeatletter
\renewcommand\paragraph{\@startsection{paragraph}{4}{\z@}%
            {3.25ex \@plus1ex \@minus.2ex}%
            {-1em}%
            {\normalfont\normalsize\bfseries\itshape}}
\makeatother
\begin{document}

\title{Interpretable Price Bounds Estimation with Shape Constraints in Price Optimization 
}


\author{Shunnosuke Ikeda \and
        Naoki Nishimura \and
        Shunji Umetani 
}

\institute{Shunnosuke Ikeda \at 
            Recruit Co., Ltd.
            1–9–2, Chiyoda–ku, 100–6640, Tokyo, Japan \\
              \email{sns\_ikeda@r.recruit.co.jp}           
           \and
           Naoki Nishimura \at
            Recruit Co., Ltd.
            1–9–2, Chiyoda–ku, 100–6640, Tokyo, Japan \\
            \email{nishimura@r.recruit.co.jp}           
            \and
            Shunji Umetani \at
            Recruit Co., Ltd.
            1–9–2, Chiyoda–ku, 100–6640, Tokyo, Japan \\
            \email{shunji\_umetani@r.recruit.co.jp}           
}

\maketitle

\begin{abstract}
This study addresses the interpretable estimation of price bounds in the context of price optimization.
In recent years, price-optimization methods have become indispensable for maximizing revenue and profits.
However, effective application of these methods to real-world pricing operations remains a significant challenge.
It is crucial for operators responsible for setting prices to utilize reasonable price bounds that are not only interpretable but also acceptable.
Despite this necessity, most studies assume that price bounds are given constant values, and few have explored reasonable determinations of these bounds.
Therefore, we propose a comprehensive framework for determining price bounds that includes both the estimation and adjustment of these bounds.
Specifically, we first estimate price bounds using three distinct approaches based on historical pricing data.
Then, we adjust the estimated price bounds by solving an optimization problem that incorporates shape constraints.
This method allows the implementation of price optimization under practical and reasonable price bounds suitable for real-world applications.
We report the effectiveness of our proposed method through numerical experiments using historical pricing data from actual services.

\end{abstract}

\section{Introduction} \label{sec:intro}
\subsection{Background}
Product and service prices have a significant impact on consumer demand, allowing companies to increase their revenue and profits through appropriate pricing strategies.
Moreover, rapid advancements in information technology and e-commerce have facilitated the integration of real-time consumer demand data into pricing strategies.
This integration increases the impact of the pricing strategies on revenue and profits.
Consequently, the significance of data-driven price optimization is being increasingly highlighted.

Despite these advancements, many companies continue to rely on the intuition and experience of pricing operators to set the prices of products and services.
Manual pricing by operators incurs labor costs and is inherently limited in its ability to maximize revenue and profits.
Consequently, there is a growing trend towards the development of automated pricing systems that optimize revenue and profit by incorporating analyses of customer and competitor responses, sales performance, and pricing trends.
Investment in these systems and tools has become particularly prevalent in the retail industry 
\cite{levy2004emerging}.

However, the implementation and practical application of these systems in actual operations are not straightforward.
In practice, although the optimal price suggested by the system should ideally be adopted as the product price, operators must still take responsibility for setting the prices.
For example, operators may find it difficult to trust a system if the suggested prices deviate significantly from the prices they previously set.
Therefore, to support effective decision-making, these systems must propose prices within a range deemed acceptable by operators.

\subsection{Related work}\label{sec:related}
Price optimization is a critical decision-making challenge for many organizations and has been studied extensively as a central topic in the field of marketing. 
Price optimization has been successfully applied across a diverse range of industries, including retail \cite{bitran1997periodic,subrahmanyan1996developing}, car rental \cite{carroll1995evolutionary,geraghty1997revenue}, hotels \cite{bitran1995application,bitran1996managing}, internet providers \cite{nair2001application}, passenger railways \cite{ciancimino1999mathematical}, cruise lines \cite{ladany1991optimal}, and electricity \cite{schweppe2013spot,smith1993linear,oren1993service}.
For more detailed insights, see also comprehensive surveys on revenue management and price optimization \cite{klein2020review,kunz2014demand,bitran2003overview}.

These price-optimization problems are typically designed to derive an optimal price that maximizes revenue or profit within a given price range (i.e., lower and upper price bounds).
While many studies assume fixed constant values for price bounds, in real business operations price bounds vary depending on the product type and current prices, requiring adaptation according to the situation.

Therefore, it is essential to set appropriate price bounds for each product.
However, to the best of our knowledge, few studies have thoroughly discussed how to determine appropriate price bounds for real-world applications.
Consequently, determining the price bounds of products based on historical pricing data remains a critical challenge in making price optimization methods more applicable.

\subsection{Our Contribution}
This study aims to establish a framework for determining the price bounds of products in price optimization using historical pricing data.
To this end, we propose an interpretable framework consisting of two phases: the estimation and adjustment of price bounds.

First, we introduce three approaches for estimating price bounds: naive rules (NR), data mining (DM) techniques, and machine learning (ML) techniques.
However, owing to the limited availability of historical data, these approaches can sometimes lead to overfitting, which complicates the interpretation of price bounds by operators. 
To address this issue, we adjust the estimated price bounds by solving an optimization problem that involves shape constraints such as monotonicity, convexity, and concavity.
This problem is formulated as a convex quadratic optimization problem, which can be easily solved using optimization solvers. 

To validate the effectiveness of our proposed framework in determining the appropriate price bounds, we conducted numerical experiments using actual historical pricing data from Recruit Co., Ltd. 
We evaluated the accuracy of two methods: one that relies solely on estimation approaches, and another that integrates mathematical optimization with shape constraints.
The results of these experiments indicate that our method not only improves the accuracy of price bound estimation by mitigating overfitting, but also provides highly interpretable price bounds.

\section{Problem Setting}
In this study, we focus on the case in which the prices of multiple products are simultaneously optimized to maximize revenue.
Let $p_m$ and $q_m$ denote the price and demand for product $m \in \mathcal{M}$, respectively.
We define the price vector as $\bm{p} = (p_m)_{m \in \mathcal{M}}$.
Demand $q_m$ for product $m$ depends on both its own price and the prices of other products; hence, it is described as a function of the price vector $\bm{p}$, denoted by $q_m(\bm{p})$.
This formulation enables us to account for the effects of substitutional and complementary goods (i.e., cross elasticity of demand).

For example, consider a simplified case with two products (i.e., $\mathcal{M} = \{1, 2\}$).
Assuming that the demand functions $q_1$ and $q_2$ are linear, they can be expressed as follows:
\begin{align}
q_1 = a_{11}p_1 + a_{12}p_2 + b_1, \quad q_2 = a_{21}p_1 + a_{22}p_2 + b_2,
\end{align}
where $a_{11}, a_{12}, a_{21}, a_{22}$ are the regression coefficients, and $b_1, b_2$ are the intercepts.
In this case, positive values of $a_{12}$ and $a_{21}$ indicate complementary goods, whereas negative values indicate substitutional goods.
The total revenue for the two products can be written as
\begin{align}
p_1 q_1 + p_2 q_2 = p_1 (a_{11}p_1 + a_{12}p_2 + b_1) + p_2 (a_{21}p_1 + a_{22}p_2 + b_2).
\end{align}

Let $p^{\text{lb}}_1, p^{\text{lb}}_2$ denote the lower price bounds and $p^{\text{ub}}_1, p^{\text{ub}}_2$ represent the upper price bounds for each product.
The simple revenue maximization problem for the two products is formulated as follows:
\begin{align}
\underset{p_1, p_2~~}{\parbox{5em}{maximize}} & p_1 (a_{11}p_1 + a_{12}p_2 + b_1) + p_2 (a_{21}p_1 + a_{22}p_2 + b_2) \\
\parbox{5em}{subject to} & p^{\text{lb}}_1 \leq p_1 \leq p^{\text{ub}}_1, \\
& p^{\text{lb}}_2 \leq p_2 \leq p^{\text{ub}}_2.
\end{align}

For a more general description, let $\mathcal{P}_m$ be the feasible region for the price $p_m$.
The multi-product price optimization problem can then be formulated as follows \cite{soon2011review,ito2017optimization,ikeda2023prescriptive}:
\begin{align}
\underset{\bm p~~~}{\parbox{5em}{maximize}} & \sum_{m \in \mathcal{M}} p_m q_m(\bm{p}), \label{eq:obj} \\
\parbox{5em}{subject to} & ~p_m \in \mathcal{P}_m, \quad \forall m \in \mathcal{M}, \label{eq:price constraints}
\end{align}
where the objective function \eqref{eq:obj} represents the total revenue, and the constraints \eqref{eq:price constraints} ensure that the price $p_m$ is selected from a feasible set of prices satisfying the operational constraints.
This formulation assumes that the cross elasticity of demand can be incorporated, whereas the feasible price region for each product remains independent.

In real-world pricing operations, various operational constraints exist, including price bounds, and markup and markdown constraints \cite{bitran2003overview}.
Integrating these constraints into an optimization model is essential for developing a practical price optimization system.
However, given the diverse range of products, it is often challenging for operators to examine these constraints thoroughly.
Operators also frequently rely on their intuition and experience in decision making, which complicates the explicit definition of these constraints within the model.
Consequently, formulating a price-optimization problem that fully accounts for all real-world operational constraints is particularly challenging.
Moreover, even when these constraints can be explicitly defined, their inclusion often makes the optimization problem significantly more complex to solve \cite{gallego2019revenue}.

To address this issue, we derive the feasible region $\mathcal{P}_m$ for the price $p_m$ using historical pricing data as a constraint on the price bounds for each current price.
In this study, we focus on determining the profit margin driven by cost-based pricing rather than setting direct prices.
Let $c_m$ represent the cost and $r_m$ represent the profit margin of product $m$, respectively.
The price of product $p_m$ can then be described as follows:
\begin{align}
p_m = (1 + r_m) \cdot c_m, \quad \forall m \in \mathcal{M}.
\end{align}
Consequently, the task of estimating the feasible region $\mathcal{P}_m$ for price $p_m$ shifts to determining the feasible region $\mathcal{R}_m$ for the profit margin $r_m$.

\begin{figure}[htbp]
\centering
  \begin{minipage}[b]{0.32\linewidth}
    \centering
    \includegraphics[keepaspectratio, scale=0.4]{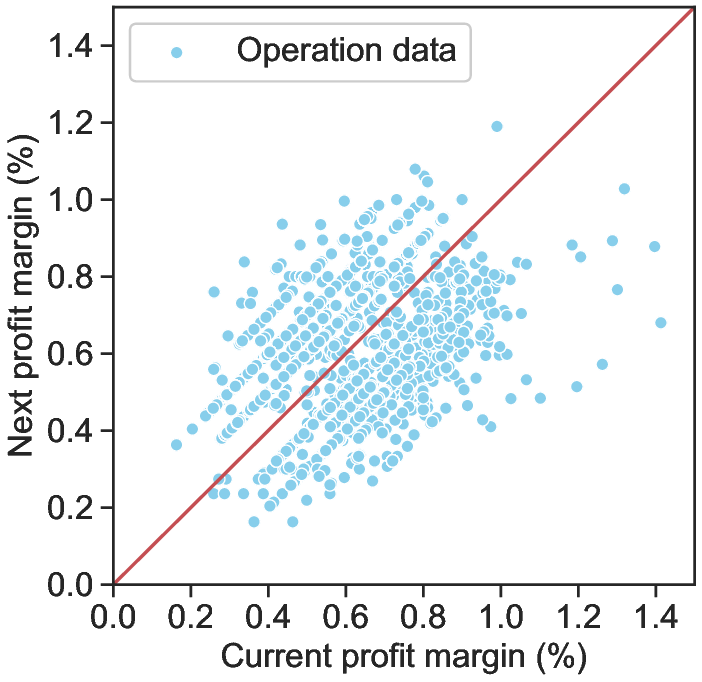}
    \subcaption{Product A}
  \end{minipage}
  \begin{minipage}[b]{0.32\linewidth}
    \centering
    \includegraphics[keepaspectratio, scale=0.4]{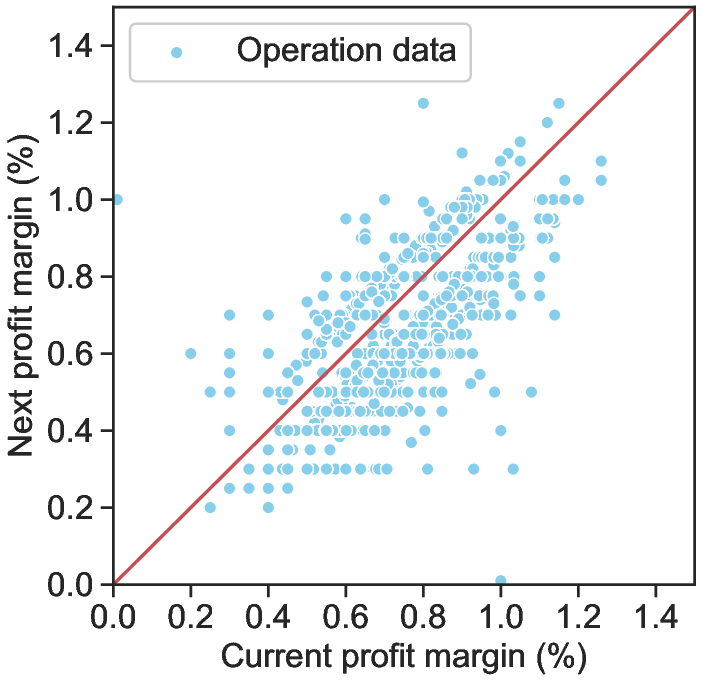}
    \subcaption{Product B}
  \end{minipage}
  \begin{minipage}[b]{0.32\linewidth}
    \centering
    \includegraphics[keepaspectratio, scale=0.4]{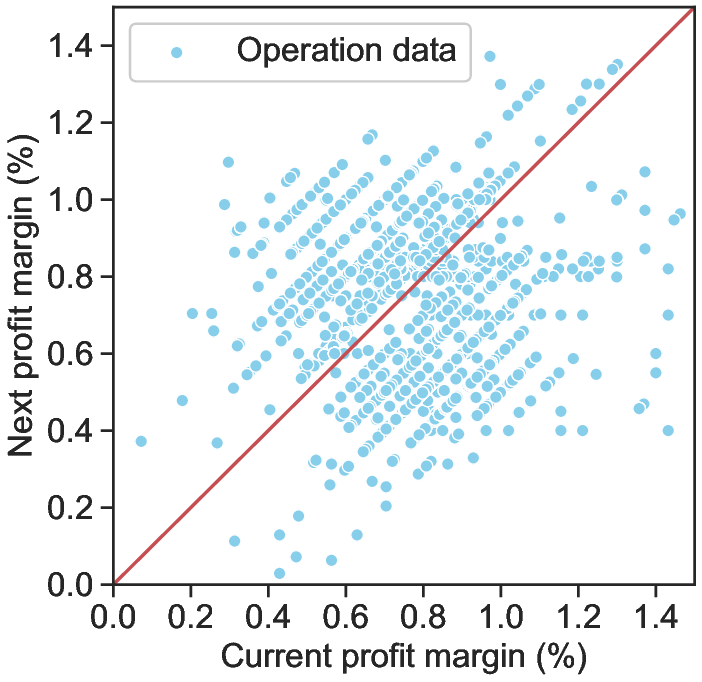}
    \subcaption{Product C}
  \end{minipage}
  \caption{Real historical pricing data for three products} \label{fig:price_log}
\end{figure}

Fig.~\ref{fig:price_log} shows the real historical pricing data for three products scaled by a uniform value.
The data in the lower-right triangle (the area below the red diagonal line) indicate a decrease in prices, whereas the data in the upper-left triangle indicate an increase.
This figure also illustrates that the operating range of the profit margin strongly depends on the current profit margin.
Moreover, it reveals that the operating ranges of the profit margin differ across various product types, even when the current profit margins are identical.

\begin{figure}[htbp]
\centering
\begin{minipage}[b]{0.32\linewidth}
\centering
\includegraphics[keepaspectratio, scale=0.18]{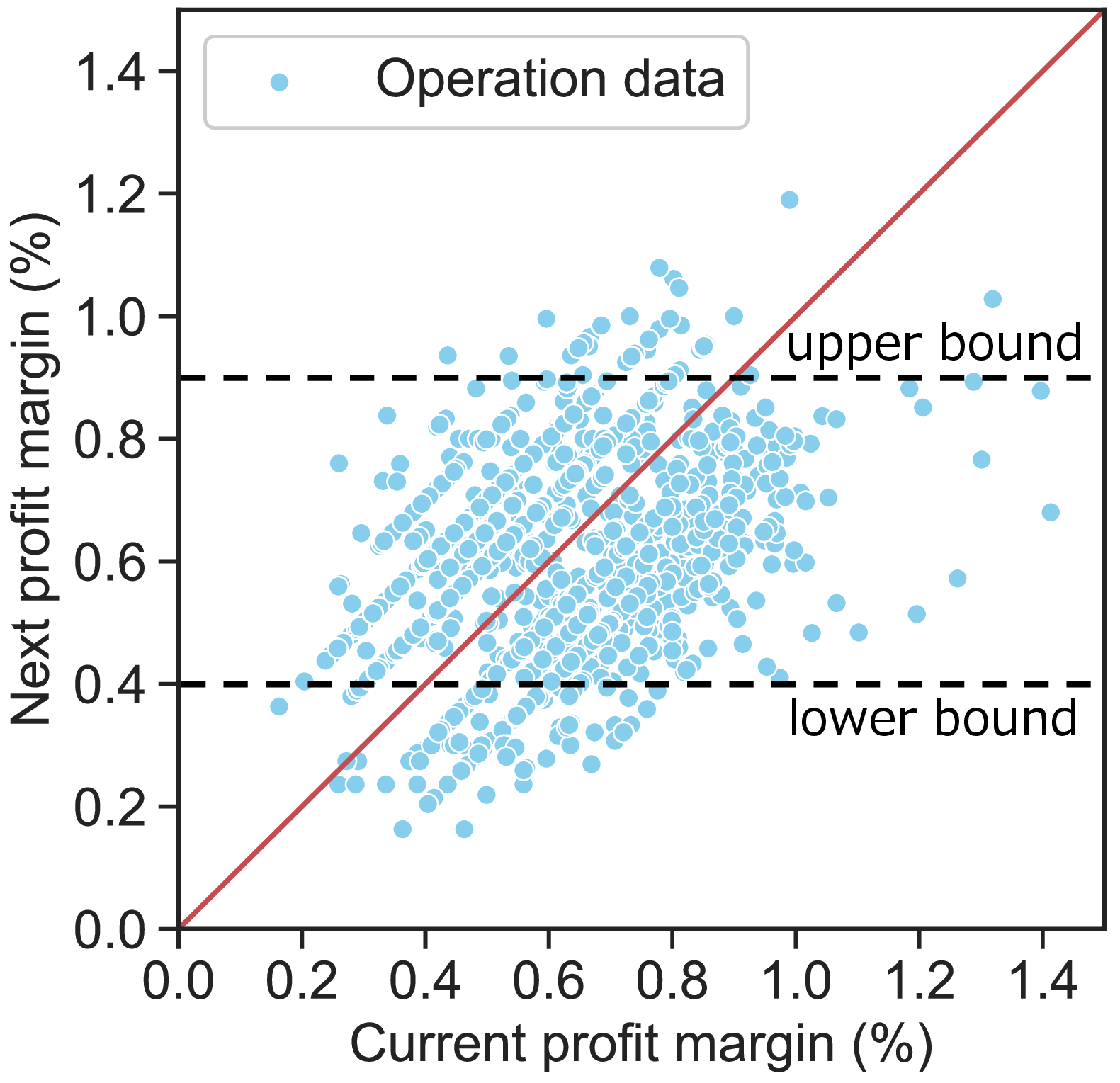}
\subcaption{Based on FCV (narrow)}
\end{minipage}
\begin{minipage}[b]{0.32\linewidth}
\centering
\includegraphics[keepaspectratio, scale=0.18]{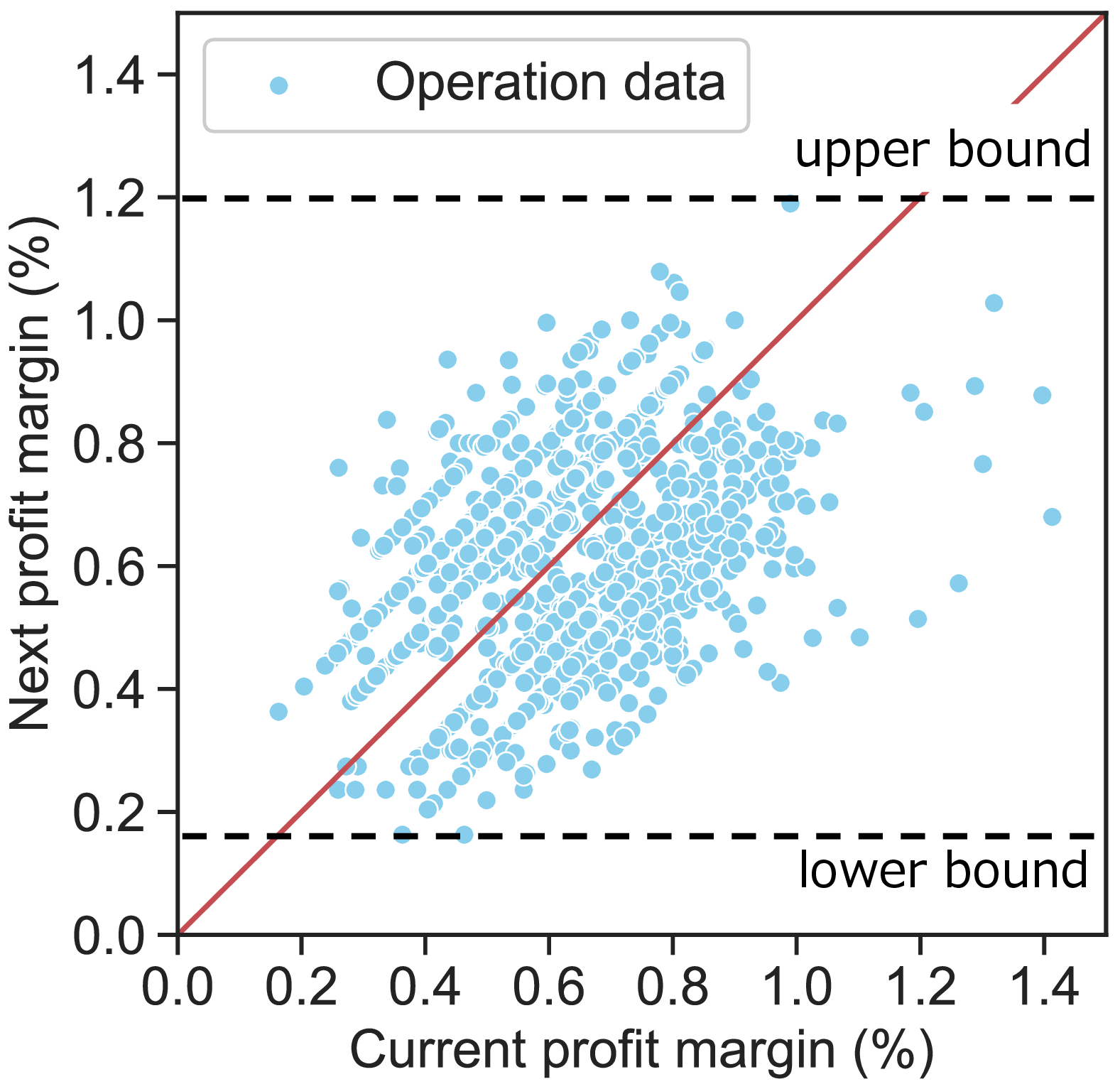}
\subcaption{Based on FCV (wide)}
\end{minipage}
\begin{minipage}[b]{0.32\linewidth}
\centering
\includegraphics[keepaspectratio, scale=0.18]{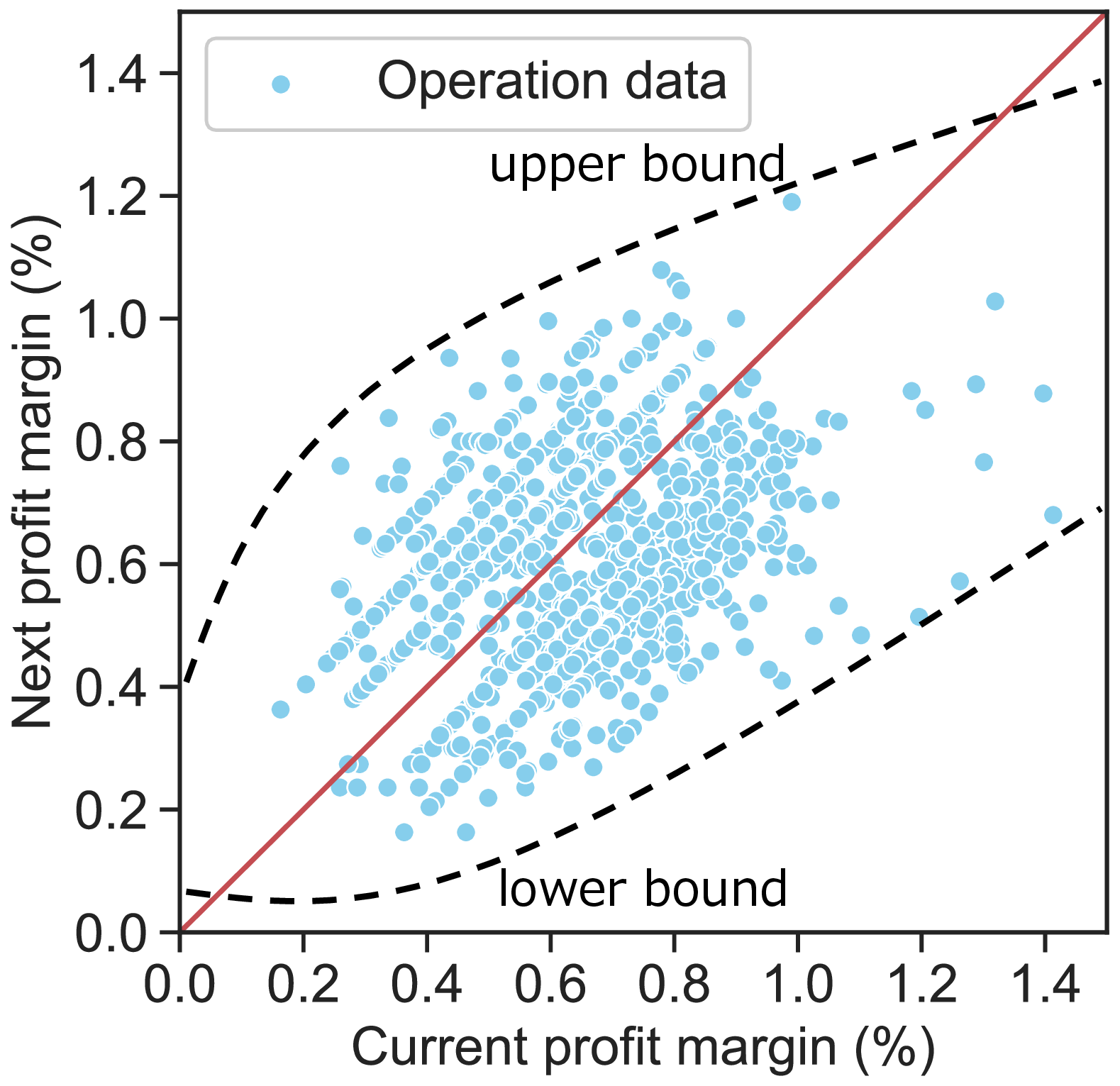}
\subcaption{Based on HPD}
\end{minipage}
\caption{Limitation and possibility of profit margin bounds based on fixed constant values (FCV) and historical pricing data (HPD)} \label{fig:limitation}
\end{figure}

However, as discussed in Section \ref{sec:related}, most previous studies assume fixed constant values for price bounds.
To illustrate the limitations of this approach and demonstrate how these limitation can be overcome using historical data, Fig. \ref{fig:limitation} presents examples of profit margin bounds for Product A, based on both fixed constant values and historical pricing data.
Figs. \ref{fig:limitation}(a) and \ref{fig:limitation}(b) depict narrow and wide bounds based on fixed constant values, respectively.
These figures show that narrow bounds can lead to lost pricing opportunities, whereas wide bounds can result in unrealistic pricing decisions.
Fig. \ref{fig:limitation}(c) illustrates bounds derived from historical pricing data, demonstrating that these bounds are appropriately adjusted for each current profit margin.

Therefore, we estimate the profit margin bounds based on the current profit margin.
We first discretize the domain of the profit margin $[r^{\min}, r^{\max}]$ using a specified step size $\delta$.
Let $i \in \mathcal{I}$ denote an index corresponding to the current profit margin, with $\mathcal{I}$ defined as $\{1, 2, \dots, \lceil (r^{\max} - r^{\min}) / \delta \rceil\}$.
We define $r^{\text{lb}}_{im}$ and $r^{\text{ub}}_{im}$ as the lower and upper bounds of the profit margin, respectively, for indices $i$ and $m$.
For index $i$ and product $m$, the price-bound constraint is equivalent to the following constraint:
\begin{align}
r^{\text{lb}}_{im} \leq r_m \leq r^{\text{ub}}_{im}.
\end{align}

In this problem setting, we assume that the profit margin bounds depend solely on the current profit margin.
Although it is possible to derive price bounds that consider a series of past pricing movements, such bounds are structurally more complex, making them more difficult to visualize and interpret.
Because the goal of this study is to derive highly interpretable price bounds, we structured the problem such that the profit margin bounds depend only on the current profit margin.
Thus, this study focuses on determining the appropriate bounds for the profit margin (i.e., $r^{\text{lb}}_{im}$ and $r^{\text{ub}}_{im}$), using historical pricing data.
In the subsequent discussions, the subscripts denoting product $m$ are omitted because the profit margin are estimated individually for each product $m$.

\section{Our Framework for Interpretable Price Bounds Estimation}
In this section, we propose a framework for determining the appropriate bounds of the profit margin by incorporating both the estimation and adjustment processes.
First, we present three estimation approaches based on naïve rules, DM techniques, and ML techniques.
Subsequently, we detail the adjustment of the estimated bounds using mathematical optimization.

\subsection{Price Bounds Estimation} \label{sec:pbe}
While there are many alternatives for estimating price bounds from two-dimensional pricing data, here we present three representative estimation approaches: NR-based, DM-based, and ML-based approaches.

\paragraph{NR-Based Approach}
First, we estimate the profit margin bounds using quantiles.
Quantiles are values that divide a finite set of data into $q$ subsets of approximately equal sizes, ensuring that each subset contains approximately the same amount of data.

Let $\mathcal{R}_i$ denote the set of next profit margins in the historical pricing data corresponding to index $i$.
We then sort the elements of $\mathcal{R}_i$ in ascending order as $\bar{r}_{i1} \leq \bar{r}_{i2} \leq \cdots \leq \bar{r}_{i|\mathcal{R}_i|}$.
For a real number $q \in [0, 1]$, the $q$-quantile $r^{\text{qnt}}_{i}(q)$ of $\mathcal{R}_i$ is defined as
\begin{align}
t &= 1 - q + q|\mathcal{R}_i|, \\
r_{i}^{\text{qnt}}(q) &= \begin{cases}\bar{r}_{it}, & \text { if } t \in \mathbb{N}, \\ (\lceil t\rceil-t) \bar{r}_{i\lfloor t\rfloor}+(t-\lfloor t\rfloor) \bar{r}_{i\lceil t\rceil}, & \text { otherwise }. \end{cases}
\end{align}

\begin{figure}[H]
\centering
\includegraphics[width=0.75\textwidth]{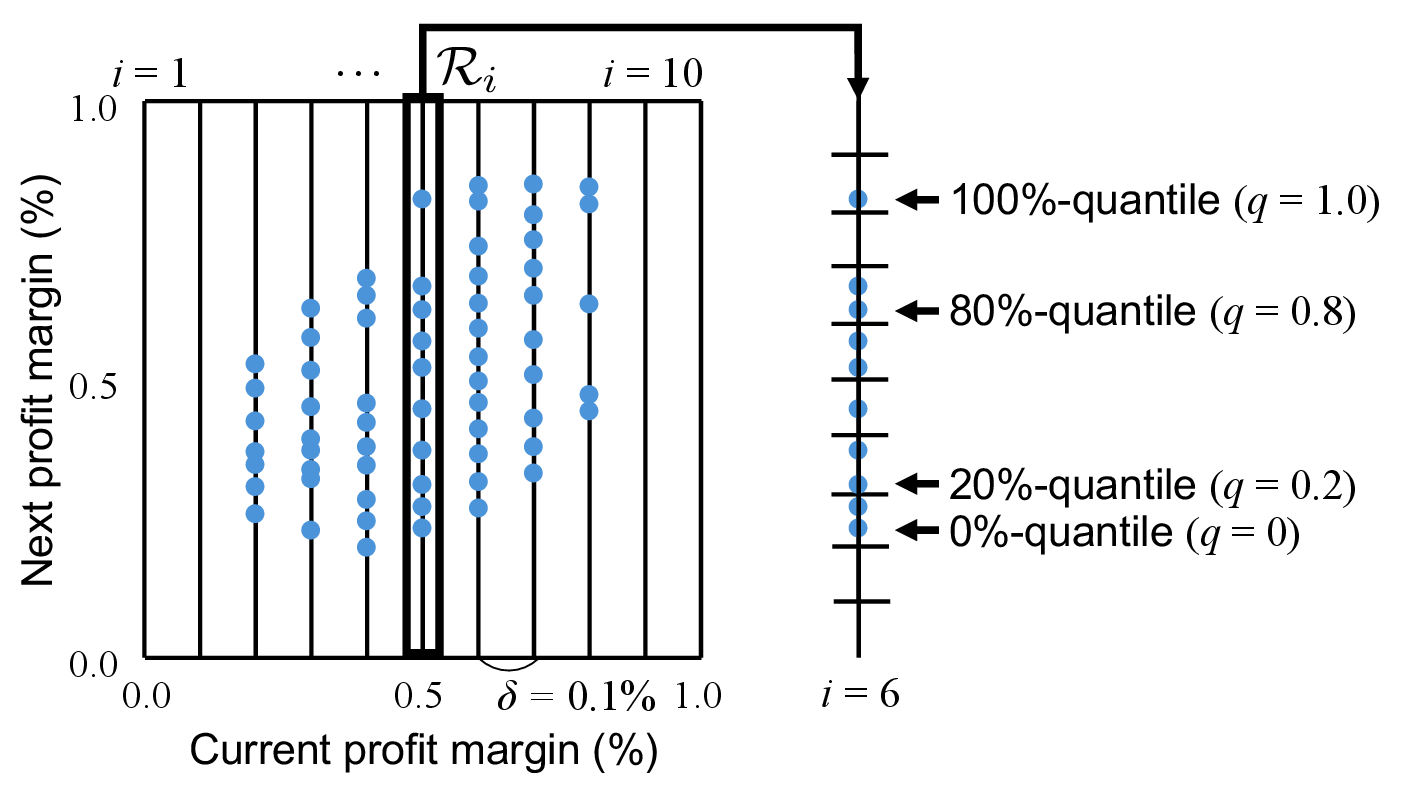}
\caption{Example of estimated bounds using quantiles}\label{fig:example_qnt}
\end{figure}

Fig.~\ref{fig:example_qnt} shows an example of estimating the profit margin bounds using quantiles.
In this approach, the $q$-quantile is designated as the estimated lower bound $\hat{r}^{\text{lb}}_{i}$ and the $1-q$ quantile as the estimated upper bound $\hat{r}^{\text{ub}}_{i}$, as follows:
\begin{align}
\hat{r}^{\text{lb}}_{i} = r_{i}^{\text{qnt}}(q), \quad \hat{r}^{\text{ub}}_{i} = r_{i}^{\text{qnt}}(1-q), \quad \forall i \in \mathcal{I}.
\end{align}

\paragraph{\textbf{DM-Based Approach}}
Next, we use association rules to estimate the profit margin bounds.
Association rule mining, proposed by Agrawal et al. \cite{agrawal1993mining}, is a fundamental technique in data mining.
This method excels at extracting rules from large datasets and has been extensively applied across various fields.
For instance, it has been used to analyze the relationships between products, as detailed in several comprehensive surveys \cite{karthikeyan2014survey,zhang2010survey,hipp2000algorithms}.

Specifically, we use a single association rule for one-dimensional numerical data to estimate the profit margin bounds (i.e., $r^{\text{lb}}_{i}$ and $r^{\text{ub}}_{i}$) for each index $i$.
The performance of the association rules is evaluated using two critical metrics: \emph{support} and \emph{confidence}. 

Analogous to index $i$, let $j \in \mathcal{J}$ denote the index of the profit margin that follows the current profit margin at index $i$.
In addition, we define $[s,t]$ as the index interval corresponding to the estimated bounds $\hat{r}^{\text{lb}}_{i}$ and $\hat{r}^{\text{ub}}_{i}$.
Let $v_{ij}$ be a binary value set to 1 if any operation within the profit margin occurs between indices $j$ and $j+1$ for index $i$, and 0 otherwise.

Support, denoted as $\textit{sup}_i \in [0, 1]$, and confidence, denoted as $\textit{conf}_i \in [0, 1]$, for the index interval $[s,t]$ corresponding to index $i$ of the current profit margin, are then defined as follows:
\begin{align}
    \textit{sup}_i = \frac{\sum_{j = s}^{t} v_{ij}}{\sum_{j \in \mathcal{J}} v_{ij}}, \quad \forall i \in \mathcal{I}, \\
    \textit{conf}_i = \frac{\sum_{j = s}^{t} v_{ij}}{\sum_{j = s}^{t} 1}, \quad \forall i \in \mathcal{I}. 
\end{align}

\begin{figure}[H]
\centering
\includegraphics[width=0.7\textwidth]{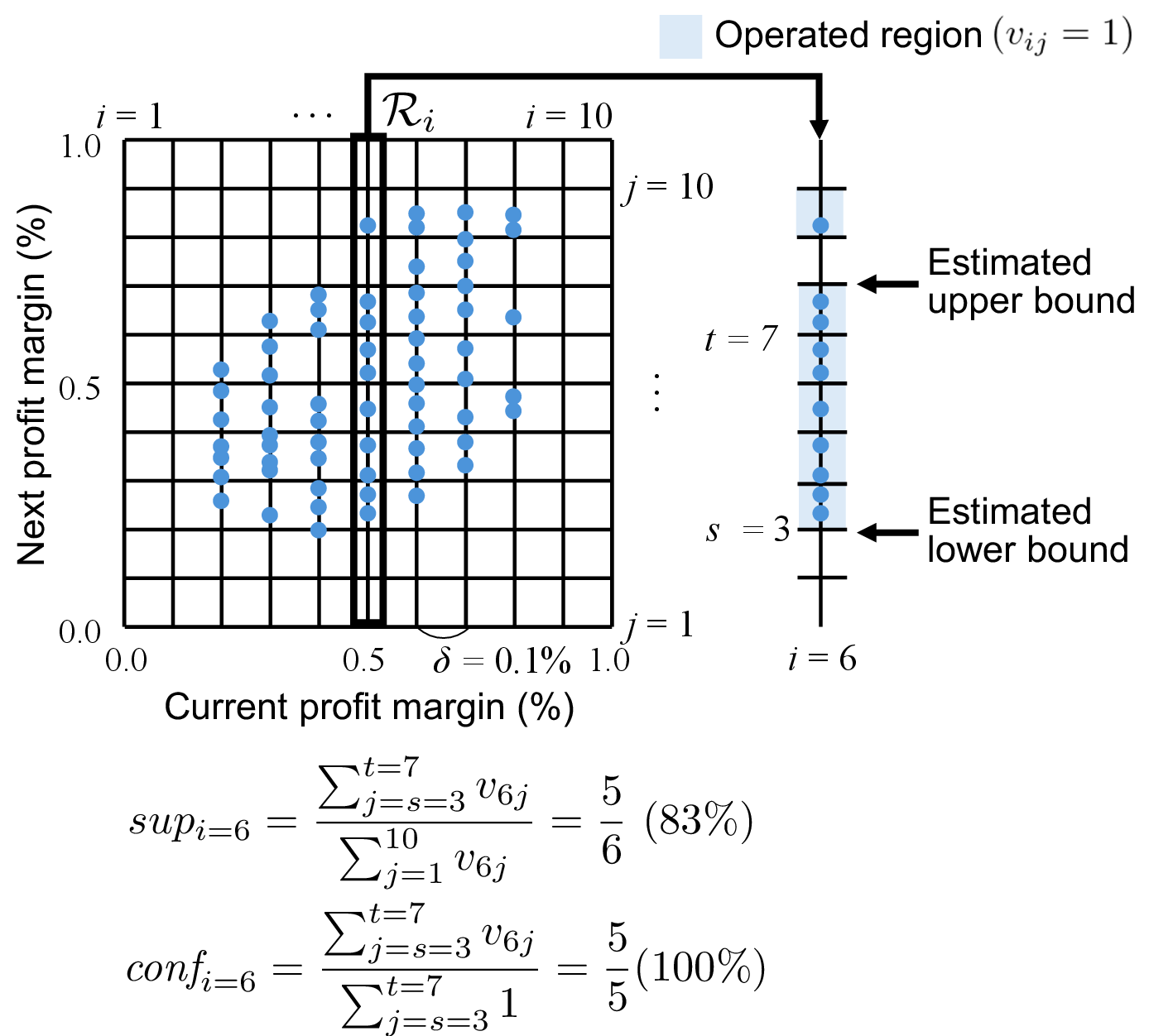}
\caption{Example of estimated bounds using association rules}\label{fig:example_dm}
\end{figure}

We determine the estimated bounds of the profit margin, $\hat{r}^{\text{lb}}_{i}$ and $\hat{r}^{\text{ub}}_{i}$, using the optimized confidence rules, which are the association rules with the highest confidence that satisfy the given minimum support. 
Fig.~\ref{fig:example_dm} illustrates an example of estimating the profit margin bounds using an optimized confidence rule, with the minimum support set to $0.8$.

\paragraph{ML-Based Approach}
Finally, we employ a one-class support vector machine (SVM) \cite{scholkopf1999support} as a machine-learning method to estimate the profit margin bounds.
One-class SVMs are primarily used for anomaly detection tasks.
This model is designed to learn the boundaries encapsulating the regions within the dataset where the samples are predominantly located.
Subsequently, it identifies data points that lie outside these predefined boundaries as anomalies.

Let $K$ denote the total number of operation data points and $k$ the index of these data points, where each operation point $\bm{x}_k$ is characterized by the coordinates $(r^{\text{cur}}_k, r^{\text{next}}_k)$, representing the current and next profit margins, respectively.
Let $\bm{w}$ represent the weight vector of the model in the feature space, and let $\phi(\bm{x}_k)$ denote the function that maps $\bm{x}_k$ to a higher-dimensional space.
Let $\rho$ be the threshold of the discriminant function and $\xi_k$ the slack variable, where $\bm{\xi} = (\xi_1, \xi_2, \ldots, \xi_K)$.
The regularization parameter is denoted as $\nu$.

The one-class SVM problem can be formulated as follows:
\begin{align}
\underset{\bm{w}, \bm{\xi}, \rho~~~}{\parbox{5em}{minimize}}  ~& \frac{1}{2}|\bm{w}|^2 + \frac{1}{\nu K} \sum_{k=1}^K \xi_k - \rho, \\
\parbox{5em}{subject to} ~& \langle \bm{w}, \phi(\bm{x}_k) \rangle \geq \rho - \xi_k, \quad k = 1, \ldots, K, \\
& \xi_k \geq 0, \quad k = 1, \ldots, K,
\end{align}
where $\langle \cdot, \cdot \rangle$ denotes the inner product.
Furthermore, for operation data $\bm{x}_k$, the anomaly score is calculated as follows:
\begin{align}
    f(\bm{x}_k)=\operatorname{sgn}(\langle\bm{w}, \phi(\bm{x}_k)\rangle -\rho),
\end{align}
where $\operatorname{sgn}(\cdot)$ is the sign function.
This function returns $+1$ (indicating normal) if the calculated score exceeds $\rho$ and $-1$ (indicating abnormal) otherwise.
The boundary is defined as the point at which the output of the function switches from $-1$ to $+1$ or vice versa.

Let $\mathcal{B}_i$ be the set of the next profit margins located on the boundary corresponding to index $i$. 
In this approach, the estimated lower bound $\hat{r}^{\text{lb}}_{i}$ and estimated upper bound $\hat{r}^{\text{ub}}_{i}$ are defined as follows:
\begin{align}
   \hat{r}^{\text{lb}}_{i} = \min \{b \mid b \in \mathcal{B}_i\}, \quad \hat{r}^{\text{ub}}_{i} = \max \{b \mid b \in \mathcal{B}_i\}, \quad \forall i \in \mathcal{I}. 
\end{align}

\subsection{Price Bounds Adjustment}
\begin{figure}[htbp]
\centering
  \begin{minipage}[b]{0.49\linewidth}
    \centering
    \includegraphics[keepaspectratio, scale=0.23]{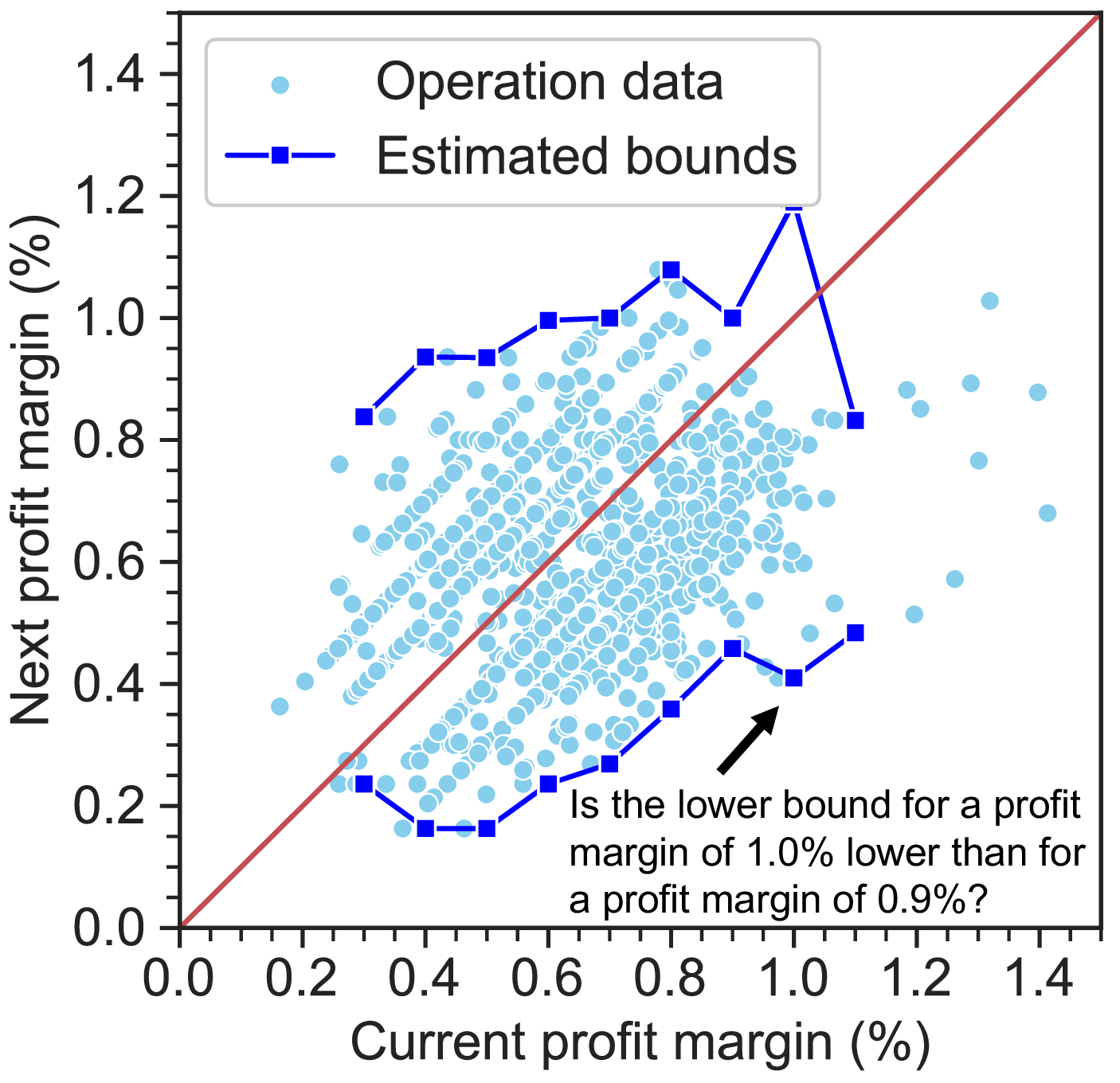}
    \subcaption{Step size: 0.1\% (i.e.,~$\delta=0.001$)~~~~~~}
  \end{minipage}
  \begin{minipage}[b]{0.49\linewidth}
    \centering
    \includegraphics[keepaspectratio, scale=0.23]{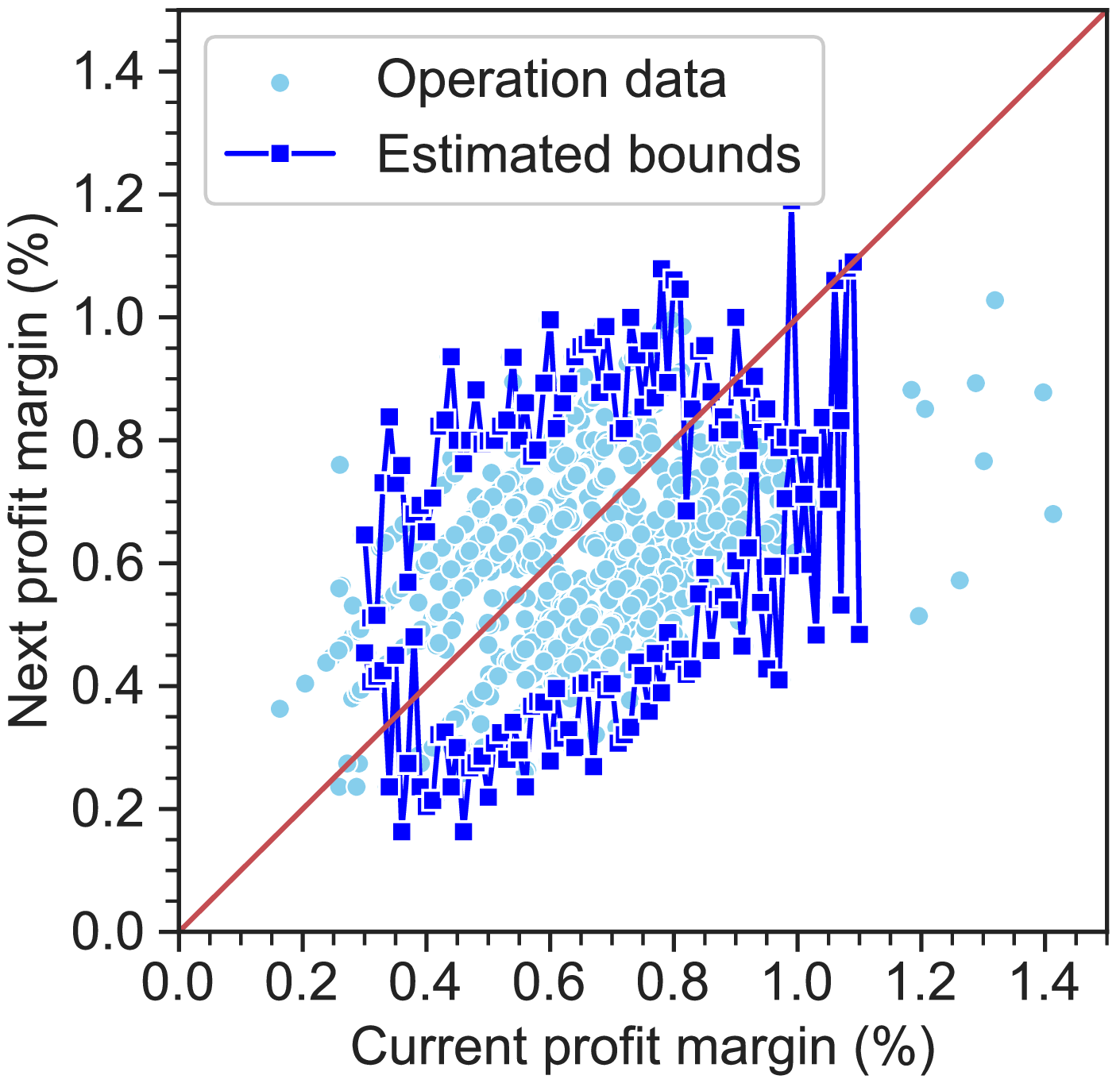}
    \subcaption{Step size: 0.01\% (i.e.,~$\delta=0.0001$)}
  \end{minipage}
  \caption{Estimated bounds of the profit margin using the DM-based approach (for the minimum support of 1.0)} \label{fig:acc_result}
\end{figure}

Fig.~\ref{fig:acc_result} illustrates the estimated bounds derived by using the association rules as an example.
Although these bounds cover most of the historical operations, several issues remain.

As shown in the left figure, the lower bound for the historical profit margin of 1.0\% is smaller than that of 0.9\%, resulting in an inconsistent trend in the estimated bounds, which may be difficult for operators to interpret.
Furthermore, the sample size varies significantly across profit margins, which greatly affects the accuracy of the estimated bounds.

In the right-hand figure, where the step size $\delta$ of the profit margin is set to 0.01\%, the bounds of the profit margin exhibit high volatility and multiple extreme values.
Such volatility and extremes deviate from operators' expectations, making it difficult to identify a clear and consistent trend, thus challenging the effective use of pricing systems.

To address these issues, we employ mathematical optimization to adjust the estimated bounds derived using the aforementioned approaches.
During this process, we introduce shape constraints into the optimization problem to reflect the characteristics of the pricing operations.
The proposed method builds on our previous work \cite{iwanaga2016estimating,nishimura2018latent,iwanaga2019improving,nishimura2023predicting} in which shape constraints based on prior knowledge were used to estimate product choice probabilities on e-commerce platforms.

We introduce three shape constraints to reflect the following pricing operation properties:
\begin{itemize}
\item \textbf{Monotonicity}: 
Both the lower and upper bounds of the profit margin are adjusted to increase monotonically as the current profit margin rises.
\vspace{1mm}
\item \textbf{Convexity of Lower Bounds}: 
The rate of increase in the lower bound of the profit margin accelerates as the current profit margin rises.
\vspace{1mm}
\item \textbf{Concavity of Upper Bounds}: 
The rate of increase in the upper bound of the profit margin decelerates as the current profit margin rises.
\end{itemize}

Let $\hat{r}_i^{\text{lb}}$ and $\hat{r}_i^{\text{ub}}$ denote the estimated lower and upper bounds of the profit margin, respectively, obtained using the estimation approaches described in Section \ref{sec:pbe}.
Additionally, let $w_i$ represent the weight coefficient for index $i$, defined in this study as the number of data points for index $i$, and let $r_i^{\text{lb}}$ and $r_i^{\text{ub}}$ denote the adjusted lower and upper bounds, respectively.
The vectors of these adjusted bounds are defined as $\bm{r}^{\text{lb}} = (r_i^{\text{lb}})_{i \in \mathcal{I}}$ and $\bm{r}^{\text{ub}} = (r_i^{\text{ub}})_{i \in \mathcal{I}}$.

The optimization problem for adjusting the estimated bounds can be formulated as a convex quadratic optimization problem incorporating the aforementioned properties of pricing operations, as follows:
\begin{align}
\underset{\bm{r}^{\text{lb}}, \bm{r}^{\text{ub}}}{\parbox{5em}{minimize}} ~&\sum_{i \in \mathcal{I}} w_i\Bigl\{(r_i^{\text{lb}} - \hat{r}_i^{\text{lb}})^2+(r_i^{\text{ub}} - \hat{r}_i^{\text{ub}})^2 \Bigr\}, \label{eq:obj2} \\
\parbox{5em}{subject to} ~& r_{i-1}^{\text{lb}} \leq r_i^{\text{lb}} , \quad \forall i \in \mathcal{I} \backslash\{1\}, \label{eq:monotone_l} \\
& r_{i-1}^{\text{ub}} \leq r_i^{\text{ub}}, \quad \forall i \in \mathcal{I} \backslash\{1\}, \label{eq:monotone_u}\\
& r_i^{\text{lb}} - r_{i-1}^{\text{lb}} \geq r_{i-1}^{\text{lb}} - r_{i-2}^{\text{lb}}, \quad \forall i \in \mathcal{I} \backslash\{1,2\}, \label{eq:convex_l} \\
& r_i^{\text{ub}} - r_{i-1}^{\text{ub}} \leq r_{i-1}^{\text{ub}} - r_{i-2}^{\text{ub}}, \quad \forall i \in \mathcal{I} \backslash\{1,2\}, \label{eq:concave_u}
 \\
& r_i^{\text{lb}} \leq r_i^{\text{ub}}, \quad \forall i \in \mathcal{I} \label{eq:u_l},
\end{align}
where the objective function \eqref{eq:obj2} represents the weighted sum of the squares of the differences between the estimated and the adjusted lower and upper bounds of the profit margin.
The constraints \eqref{eq:monotone_l} and \eqref{eq:monotone_u} ensure the monotonicity of the lower and upper bounds, respectively.
Constraints \eqref{eq:convex_l} and \eqref{eq:concave_u} impose convexity on the lower bounds and concavity on the upper bounds.
Finally, the constraint \eqref{eq:u_l} guarantees that the lower bound remains below the upper bound.

\section{Numerical Experiments}
In this section, we evaluate the effectiveness of our framework for estimating interpretable price bounds through numerical experiments using historical pricing data from real services.

\subsection{Experimental Design}
In our experiments, we used real pricing data for the three main products offered by Recruit Co., Ltd. during the period from January 1, 2019, to December 31, 2019.
In this study, these products are referred to as Products A, B, and C.
We also set the minimum and maximum values of the current profit margin to 0.3\% and 1.1\%, respectively (i.e., $r^{\text{min}} = 0.003, r^{\text{max}} = 0.011$).
The domain $[r^{\text{min}}, r^{\text{max}}]$ of the current profit margin covers more than $97\%$ of all operations.

To evaluate the proposed method, we used cross validation by dividing the data into five folds based on the operation date for each pricing operation.
Subsequently, for each estimation approach (i.e., NR- , DM-, and ML-based approaches), we compared the Root Mean Squared Error (RMSE) between the estimated values before and after adjustment using the four different methods:
\vspace{2mm}
\begin{description}
\item[\textbf{NA:}] No adjustment, i.e., price bounds estimation only;
\item[\textbf{MN:}] Monotonicity adjustment only;
\item[\textbf{CC:}] Convexity-concavity adjustment only;
\item[\textbf{MN-CC:}] Both monotonicity and convexity-concavity adjustments.
\end{description}
\vspace{2mm}

To validate the effectiveness of the proposed method, we defined the improvement rate as follows:
\begin{align}
    \text{Improvement rate~(\%)} = 100 \cdot \biggl(1 - \frac{\text{RMSE of the best method}}{\text{RMSE of NA}}\biggr).
\end{align}

Additionally, we considered two cases with step sizes of 0.1\% and 0.01\% (i.e., $\delta$ values of 0.001 and 0.0001, respectively).
The hyperparameters for each estimation approach were set as follows, with the other hyperparameters maintained at their default values:
\begin{itemize}
\item For the NR-based approach, the parameter $q$ was set to $\{0.00, 0.05, 0.10\}$.
\item For the DM-based approach, the minimum support was set to $\{0.8, 0.9, 1.0\}$.
\item For the ML-based approach, the regularization parameter $\nu$ was set to $\{0.01, 0.03, 0.05\}$.
\end{itemize}

In the NR-based approach, the quantiles were calculated using the quantile function from NumPy 1.24.2, a Python library designed for efficient numerical calculations.
In the ML-based approach, one-class SVM models were implemented using the OneClassSVM function in Scikit-learn 1.2.2 \cite{pedregosa2011scikit}, which is a comprehensive Python library for machine learning tools.
The quadratic optimization problem was solved using OSQP\footnote{\url{https://osqp.org/}}, version 0.6.1, which is a state-of-the-art solver for continuous optimization.

\subsection{Results and Discussion}
Tables~\ref{tab:rmse_rb}, \ref{tab:rmse_dm}, and \ref{tab:rmse_ml} list the estimation errors derived from the NR-based, DM-based, and ML-based approaches, respectively.
Both the NR-based and DM-based approaches exhibit similar trends, with the errors arranged in ascending order as follows: MN-CC, CC, MN, and NA.

Conversely, for the ML-based approach (Table~\ref{tab:rmse_ml}), the errors are ranked in ascending order as follows: CC, NA, MN-CC, and MN.
This ranking results from the tendency of the one-class SVM to initially generate a boundary for the pricing data that closely approximates the shape of the convex hull.

Therefore, while the appropriate shape constraints must be selected based on the estimation approach, these results indicate that, across various estimation approaches, shape constraints improve the accuracy of the price bound estimation.
It should also be noted that the improvement rates in Tables~\ref{tab:rmse_rb} and \ref{tab:rmse_dm} were calculated for MN-CC, and those in Table~\ref{tab:rmse_ml} were for CC.

\begin{table*}
\centering
  \caption{RMSE using the NR-based approach}
  \label{tab:rmse_rb}
  \begin{tabular}{cccrrrrr}
  \hline
   \shortstack{Step size} & $q$ & Product & \shortstack[r]{NA} & \shortstack[r]{MN} & \shortstack[r]{CC} & \shortstack[r]{MN-CC} & \shortstack[r]{Improvement} \\ \hline
   0.01\% & 0.00 & A & 0.193 & \textbf{0.181} & 0.184 & 0.182 & 5.7\% \\
       &     & B & 0.186 & \textbf{0.172} & 0.174 & 0.174 & 6.4\% \\
       &     & C & 0.295 & 0.274 & 0.275 & \textbf{0.273} & 7.5\% \\ \hline
   0.01\% & 0.05 & A & 0.176 & 0.159 & 0.162 & \textbf{0.159} & 9.5\% \\
       &     & B & 0.169 & 0.149 & 0.150 & \textbf{0.149} & 11.8\% \\
       &     & C & 0.284 & 0.259 & 0.261 & \textbf{0.258} & 9.1\% \\ \hline
   0.01\% & 0.10 & A & 0.173 & 0.155 & 0.158 & \textbf{0.155} & 10.7\% \\
       &     & B & 0.156 & 0.136 & 0.137 & \textbf{0.136} & 13.0\% \\
       &     & C & 0.285 & 0.252 & 0.256 & \textbf{0.252} & 11.7\% \\ \hline
   0.1\% & 0.00 & A & 0.191 & 0.194 & \textbf{0.175} & 0.179 & 6.6\% \\
       &     & B & 0.214 & 0.195 & 0.192 & \textbf{0.186} & 13.1\% \\
       &     & C & 0.289 & 0.287 & 0.276 & \textbf{0.274} & 5.2\% \\ \hline
   0.1\% & 0.05 & A & 0.150 & 0.141 & 0.132 & \textbf{0.131} & 13.2\% \\
       &     & B & 0.156 & 0.149 & \textbf{0.130} & 0.131 & 16.2\% \\
       &     & C & 0.194 & 0.193 & \textbf{0.170} & 0.175 & 10.1\% \\ \hline
   0.1\% & 0.10 & A & 0.146 & 0.134 & 0.129 & \textbf{0.125} & 14.3\% \\
       &     & B & 0.131 & 0.127 & 0.119 & \textbf{0.118} & 9.5\% \\
       &     & C & 0.193 & 0.187 & \textbf{0.167} & 0.168 & 13.1\% \\ \hline
   Average &     & & 0.199 & 0.186 & 0.180 & \textbf{0.179} & 10.0\% \\ \hline
  \end{tabular}
\end{table*}

\begingroup
\begin{table*}
\centering
  \caption{RMSE using the DM-based approach}
  \label{tab:rmse_dm}
  \begin{tabular}{cccrrrrr}
  \hline
   \shortstack{Step size} & \shortstack{Min. support} & Product & \shortstack[r]{NA} & \shortstack[r]{MN} & \shortstack[r]{CC} & \shortstack[r]{MN-CC} & \shortstack[r]{Improvement} \\ \hline
   0.01\% & 0.8 & A & 0.166 & \textbf{0.147} & 0.147 & 0.147 & 11.4\% \\
       &     & B & 0.148 & 0.136 & 0.135 & \textbf{0.135} & 9.0\% \\
       &     & C & 0.269 & 0.241 & 0.244 & \textbf{0.240} & 10.9\% \\ \hline
   0.01\% & 0.9 & A & 0.180 & 0.164 & 0.166 & \textbf{0.164} & 9.0\% \\
       &     & B & 0.178 & \textbf{0.162} & 0.163 & 0.162 & 8.6\% \\
       &     & C & 0.283 & 0.256 & 0.258 & \textbf{0.256} & 9.6\% \\ \hline
   0.01\% & 1.0 & A & 0.193 & \textbf{0.181} & 0.184 & 0.182 & 5.7\% \\
       &     & B & 0.186 & \textbf{0.172} & 0.174 & 0.174 & 6.4\% \\
       &     & C & 0.295 & 0.274 & 0.275 & \textbf{0.273} & 7.5\% \\ \hline
   0.1\% & 0.8 & A & 0.147 & 0.134 & 0.129 & \textbf{0.126} & 14.8\% \\
       &     & B & 0.143 & 0.136 & 0.131 & \textbf{0.129} & 10.0\% \\
       &     & C & 0.178 & 0.161 & 0.164 & \textbf{0.156} & 12.5\% \\ \hline
   0.1\% & 0.9 & A & 0.162 & 0.151 & 0.143 & \textbf{0.141} & 12.6\% \\
       &     & B & 0.157 & 0.157 & 0.147 & \textbf{0.146} & 7.3\% \\
       &     & C & 0.206 & 0.192 & 0.184 & \textbf{0.174} & 15.3\% \\ \hline
   0.1\% & 1.0 & A & 0.191 & 0.194 & \textbf{0.175} & 0.179 & 6.6\% \\
       &     & B & 0.214 & 0.195 & 0.192 & \textbf{0.186} & 13.1\% \\
       &     & C & 0.289 & 0.287 & 0.276 & \textbf{0.274} & 5.2\% \\ \hline
   Average &     & & 0.199 & 0.186 & 0.183 & \textbf{0.180} & 9.6\% \\ \hline
  \end{tabular}
\end{table*}
\endgroup

\begin{table*}
\centering
  \caption{RMSE using the ML-based approach}
  \label{tab:rmse_ml}
  \begin{tabular}{cccrrrrr}
  \hline
   \shortstack{Step size} & $\nu$ & Product & \shortstack[r]{NA} & \shortstack[r]{MN} & \shortstack[r]{CC} & \shortstack[r]{MN-CC} & \shortstack[r]{Improvement} \\ \hline
   0.01\% & 0.01 & A & 0.095 & 0.100 & \textbf{0.091} & 0.097 & 4.5\% \\
       &     & B & \textbf{0.119} & 0.120 & \textbf{0.119} & 0.121 & 0.0\% \\
       &     & C & \textbf{0.215} & 0.217 & \textbf{0.215} & 0.217 & 0.0\% \\ \hline
   0.01\% & 0.03 & A & 0.087 & 0.093 & \textbf{0.086} & 0.092 & 1.1\% \\
       &     & B & 0.056 & 0.062 & \textbf{0.056} & 0.061 & 0.1\% \\
       &     & C & 0.176 & 0.180 & \textbf{0.173} & 0.179 & 1.6\% \\ \hline
   0.01\% & 0.05 & A & 0.101 & 0.104 & \textbf{0.095} & 0.100 & 6.2\% \\
       &     & B & 0.062 & 0.068 & \textbf{0.062} & 0.068 & 0.1\% \\
       &     & C & 0.153 & 0.154 & \textbf{0.146} & 0.152 & 4.1\% \\ \hline
   0.1\% & 0.01 & A & 0.160 & 0.162 & \textbf{0.151} & 0.157 & 5.2\% \\
       &     & B & \textbf{0.126} & 0.129 & \textbf{0.126} & 0.129 & 0.0\% \\
       &     & C & \textbf{0.234} & 0.241 & \textbf{0.234} & 0.231 & 0.0\% \\ \hline
   0.1\% & 0.03 & A & \textbf{0.113} & 0.117 & 0.125 & 0.123 & -10.2\% \\
       &     & B & \textbf{0.091} & 0.103 & \textbf{0.091} & 0.103 & 0.0\% \\
       &     & C & 0.207 & 0.206 & \textbf{0.200} & 0.204 & 3.4\% \\ \hline
   0.1\% & 0.05 & A & \textbf{0.111} & 0.112 & 0.129 & 0.121 & -15.7\% \\
       &     & B & \textbf{0.093} & 0.108 & \textbf{0.093} & 0.108 & 0.0\% \\
       &     & C & 0.157 & 0.165 & \textbf{0.155} & 0.161 & 1.6\% \\ \hline
   Average &     & & 0.131 & 0.136 & \textbf{0.130} & 0.135 & 0.1\% \\ \hline
  \end{tabular}
\end{table*}
\newpage

\begin{figure}[H]
\centering
  \begin{minipage}[b]{0.49\linewidth}
    \centering
    \includegraphics[keepaspectratio, scale=0.50]{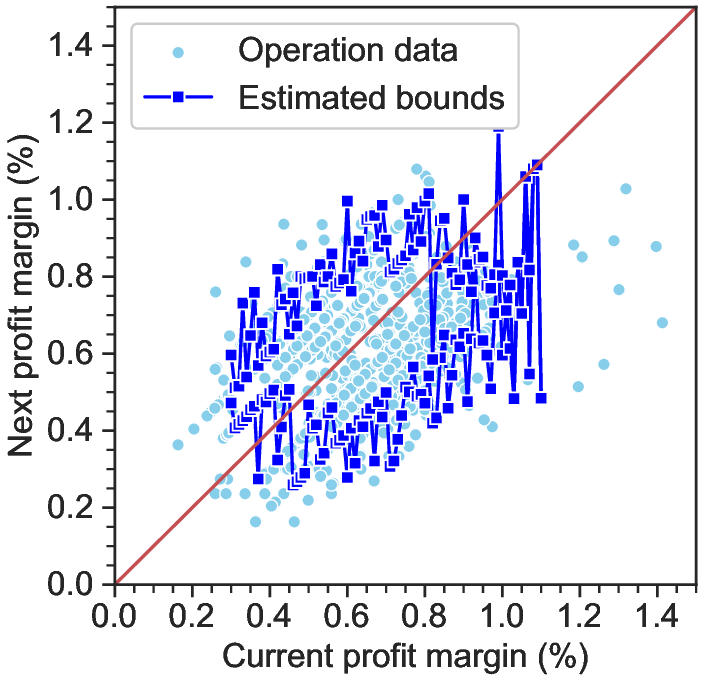}
    \subcaption{NA}
  \end{minipage}
  \begin{minipage}[b]{0.49\linewidth}
    \centering
    \includegraphics[keepaspectratio, scale=0.50]{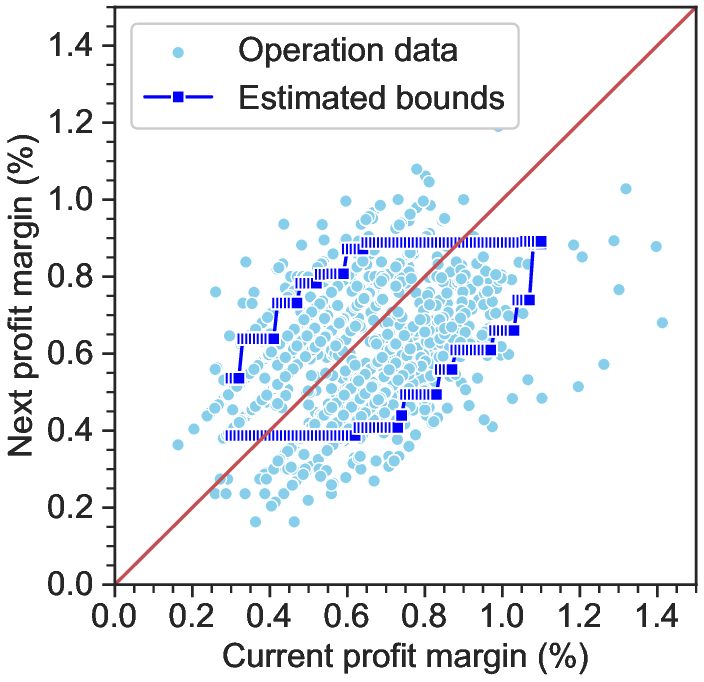}
    \subcaption{MN}
  \end{minipage}
  \begin{minipage}[b]{0.49\linewidth}
    \centering
    \includegraphics[keepaspectratio, scale=0.50]{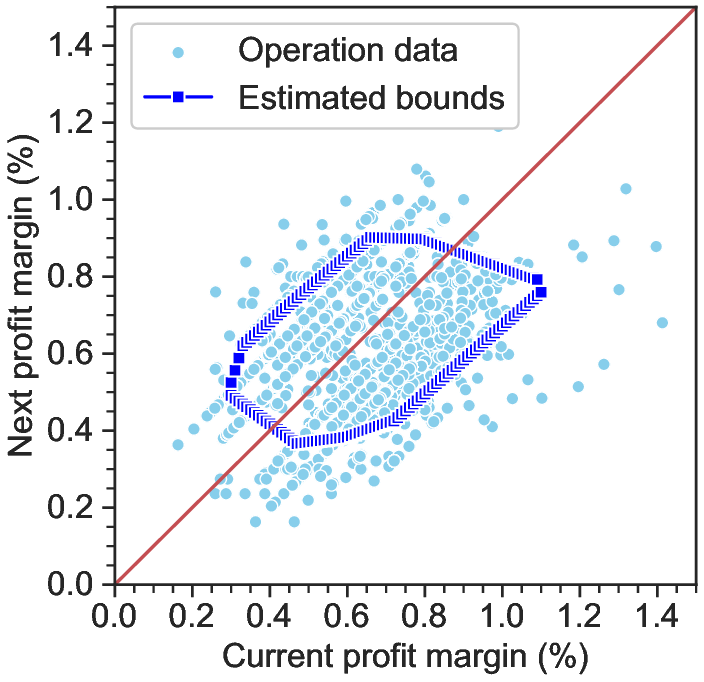}
    \subcaption{CC}
  \end{minipage}
  \begin{minipage}[b]{0.49\linewidth}
    \centering
    \includegraphics[keepaspectratio, scale=0.50]{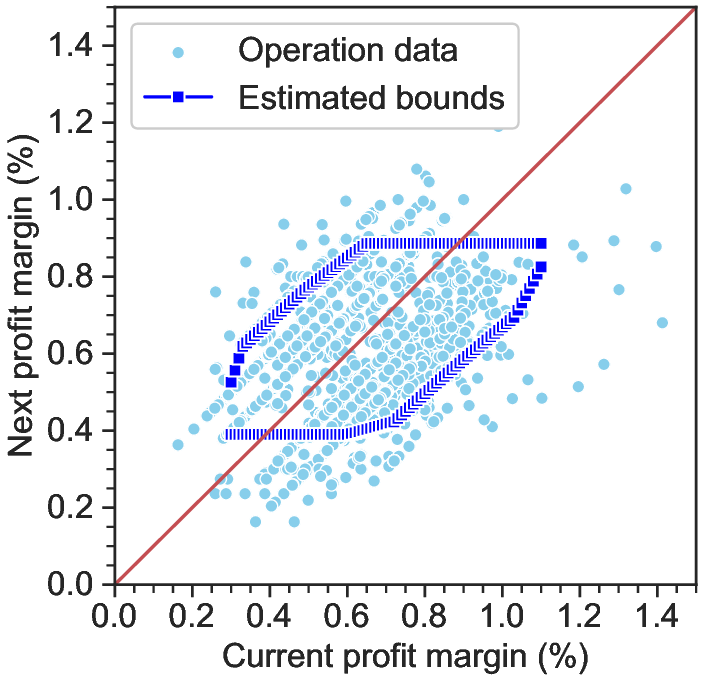}
    \subcaption{MN-CC}
  \end{minipage}
  \caption{Estimated bounds of the profit margin for Product A using the NR-based approach ($q =0.05$) with the step size of $0.01\%$} \label{fig:rb_0.01}
\end{figure}

\begin{figure}[H]
\centering
  \begin{minipage}[b]{0.49\linewidth}
    \centering
    \includegraphics[keepaspectratio, scale=0.50]{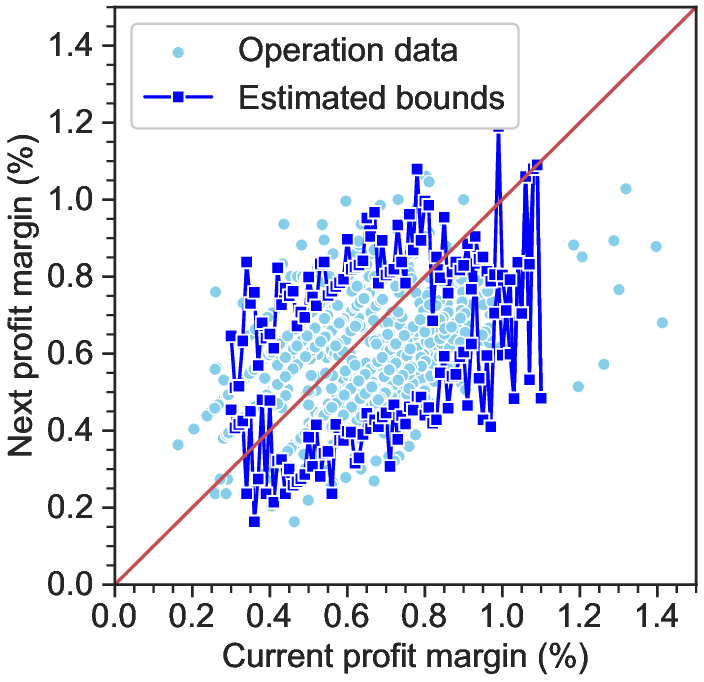}
    \subcaption{NA}
  \end{minipage}
  \begin{minipage}[b]{0.49\linewidth}
    \centering
    \includegraphics[keepaspectratio, scale=0.50]{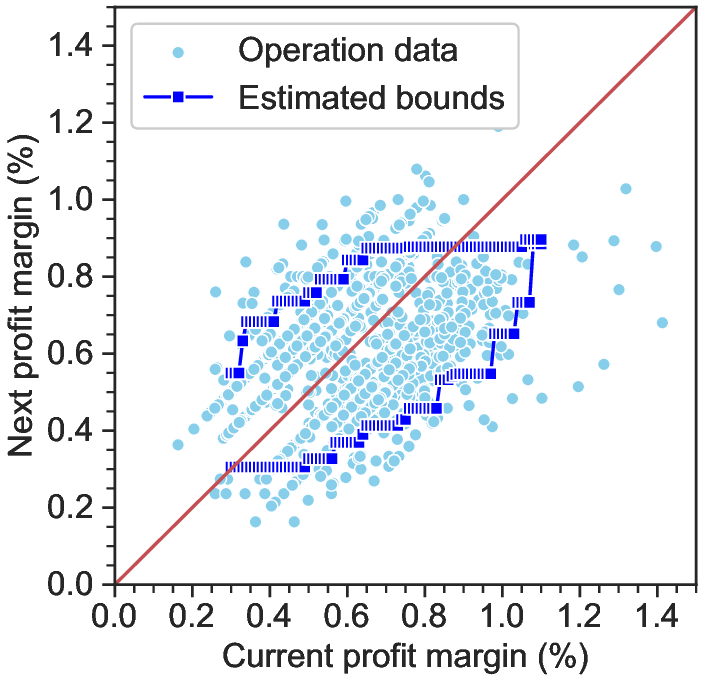}
    \subcaption{MN}
  \end{minipage}
  \begin{minipage}[b]{0.49\linewidth}
    \centering
    \includegraphics[keepaspectratio, scale=0.50]{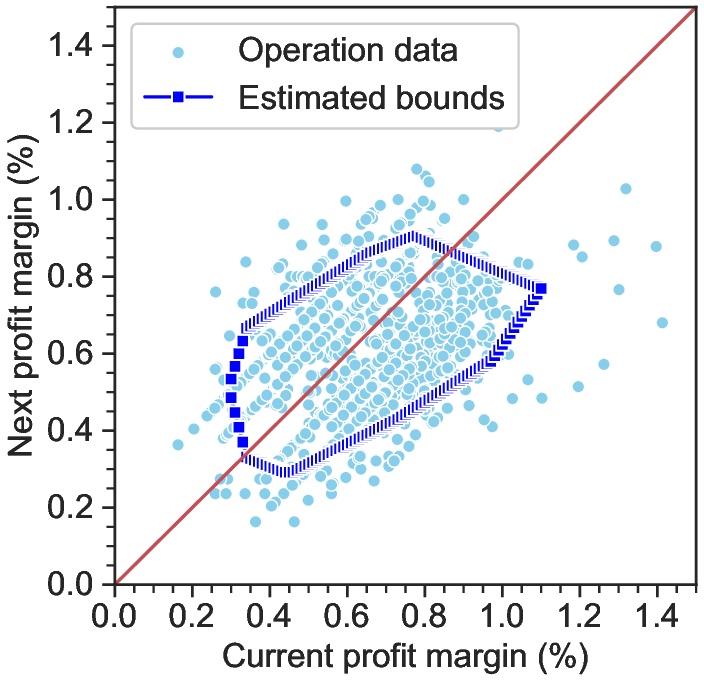}
    \subcaption{CC}
  \end{minipage}
  \begin{minipage}[b]{0.49\linewidth}
    \centering
    \includegraphics[keepaspectratio, scale=0.50]{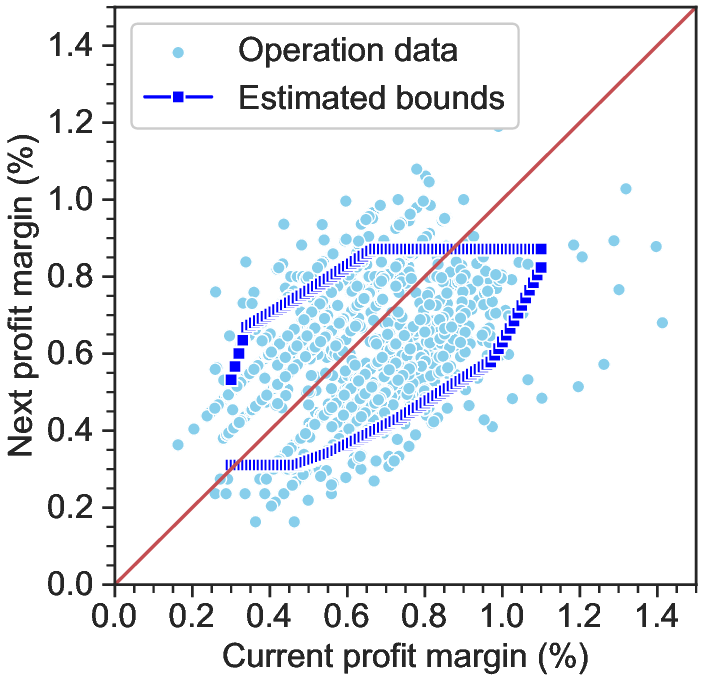}
    \subcaption{MN-CC}
  \end{minipage}
  \caption{Estimated bounds of the profit margin for Product A using the DM-based approach with the minimum support of $0.9$ and the step size of $0.01\%$} \label{fig:dm_0.01}
\end{figure}

\begin{figure}[H]
\centering
  \begin{minipage}[b]{0.49\linewidth}
    \centering
    \includegraphics[keepaspectratio, scale=0.50]{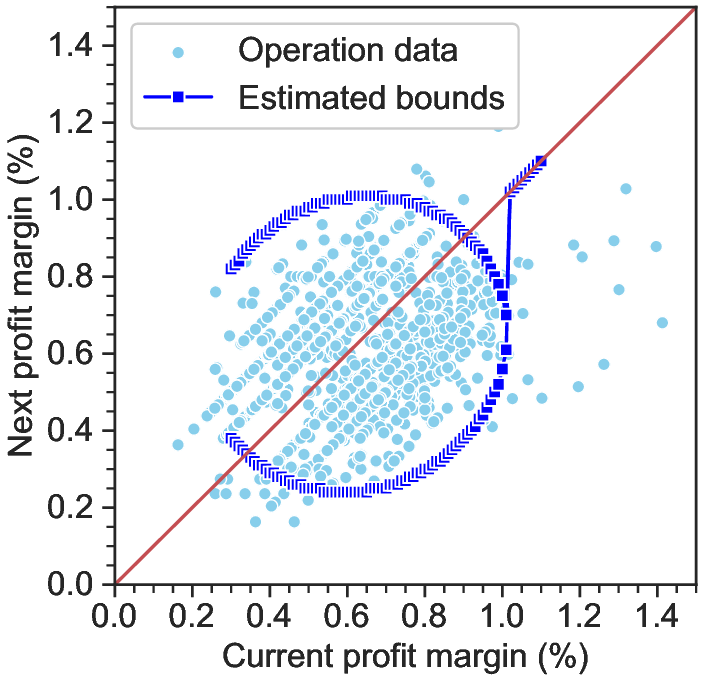}
    \subcaption{NA}
  \end{minipage}
  \begin{minipage}[b]{0.49\linewidth}
    \centering
    \includegraphics[keepaspectratio, scale=0.50]{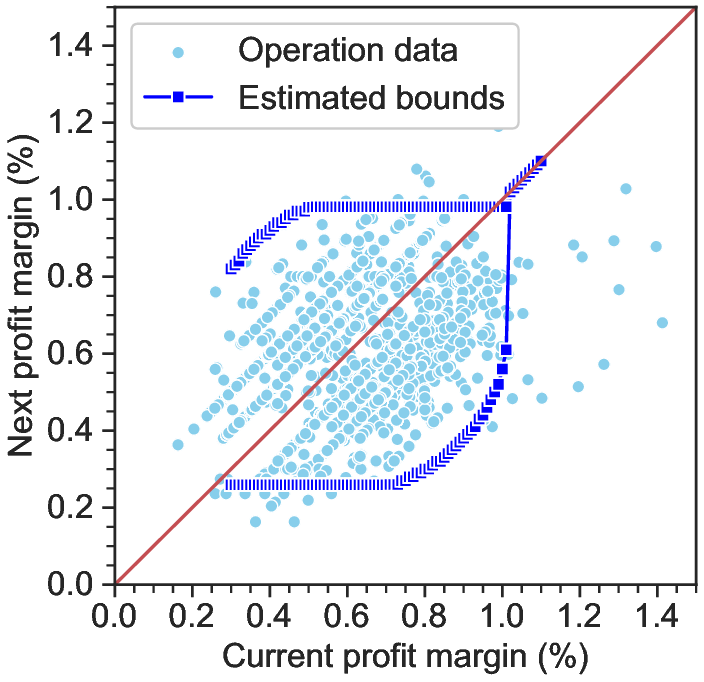}
    \subcaption{MN}
  \end{minipage}
  \begin{minipage}[b]{0.49\linewidth}
    \centering
    \includegraphics[keepaspectratio, scale=0.50]{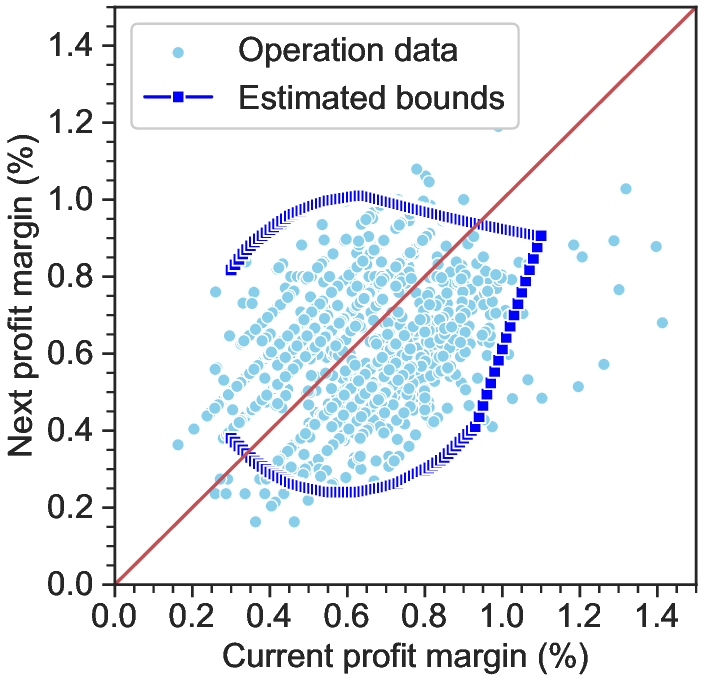}
    \subcaption{CC}
  \end{minipage}
  \begin{minipage}[b]{0.49\linewidth}
    \centering
    \includegraphics[keepaspectratio, scale=0.50]{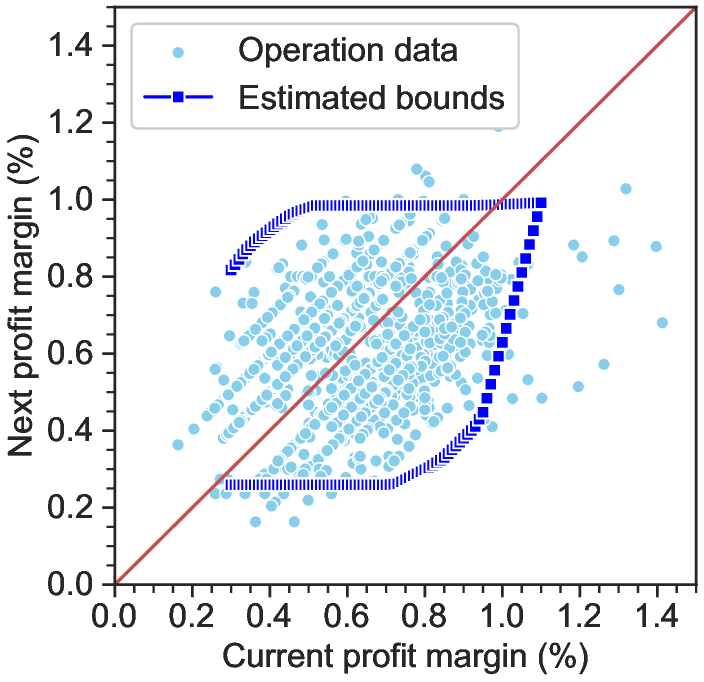}
    \subcaption{MN-CC}
  \end{minipage}
  \caption{Estimated bounds of the profit margin for Product A using the ML-based approach ($\nu =0.03$) with the step size of $0.01\%$} \label{fig:ml_0.01}
\end{figure}

Figs.~\ref{fig:rb_0.01}, \ref{fig:dm_0.01}, and \ref{fig:ml_0.01} show the estimated bounds of the profit margin for Product A derived using the NR-, DM-, and ML-based approaches, respectively, with the step size set to $0.01\%$.

In particular, when the step size was set to $0.01\%$ for both the NR- and DM-based approaches, as shown in Figs. ~\ref{fig:rb_0.01} and \ref{fig:dm_0.01}, the estimated bounds fluctuated considerably without adjustment, making it difficult for operators to interpret the trends. 
It can be observed that these estimation approaches manage to avoid overfitting through the use of shape constraints.

By contrast, the ML-based approach estimates the bounds based on all historical pricing data and is less susceptible to variation owing to smaller sample sizes.
However, even in such cases, the application of shape constraints corrects the estimated bounds to more appropriate and easily interpretable levels.

Tables \ref{tab:rmse_rb}, \ref{tab:rmse_dm}, and \ref{tab:rmse_ml} illustrate that, under both monotonicity and convexity-concavity adjustments (i.e., MN-CC), the RMSE decreases in the following order: ML-based approach $<$  NR-based approach $\approx$ DM-based approach.
The ML-based approach yielded the best results in terms of RMSE.
However, this approach tends to produce a wide range of parallel upper and lower bounds that frequently diverge from the intuition of the operator. 
Although the NR- and DM-based approaches exhibited similar behaviors, with the NR-based approach being fully practical, the DM-based approach was ultimately chosen for actual service implementation because of its closer alignment with the operator's intuition.

\begin{figure}[H]
\centering
  \begin{minipage}[b]{0.49\linewidth}
    \centering
    \includegraphics[keepaspectratio, scale=0.50]{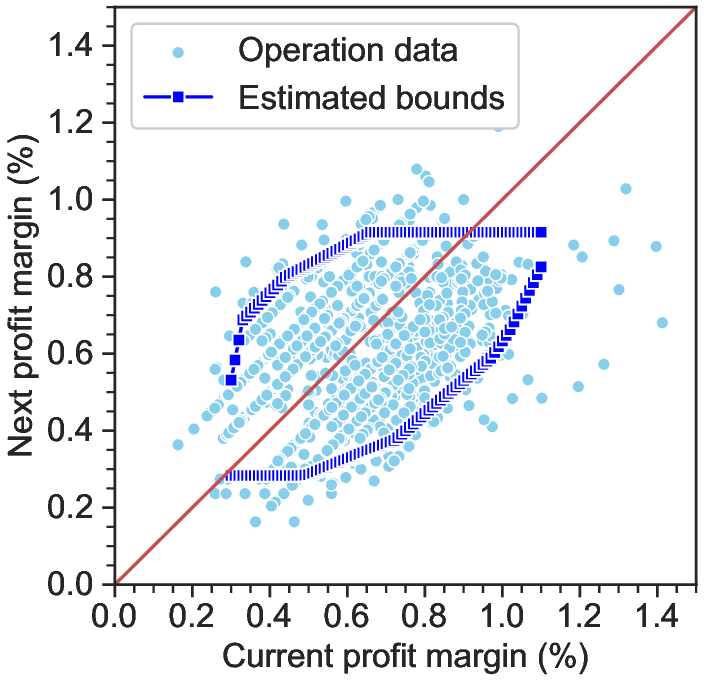}
    \subcaption{Product A}
  \end{minipage}
  \begin{minipage}[b]{0.49\linewidth}
    \centering
    \includegraphics[keepaspectratio, scale=0.50]{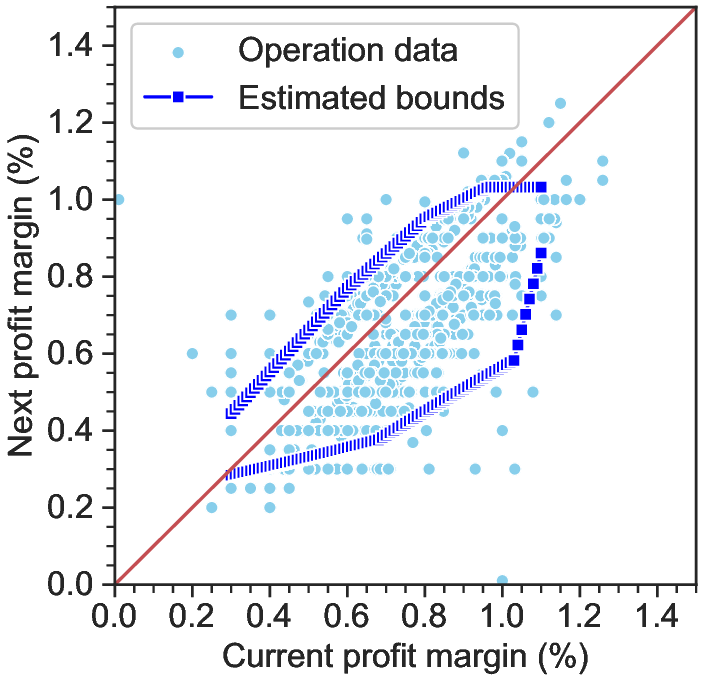}
    \subcaption{Product B}
  \end{minipage}
  \begin{minipage}[b]{0.49\linewidth}
    \centering
    \includegraphics[keepaspectratio, scale=0.50]{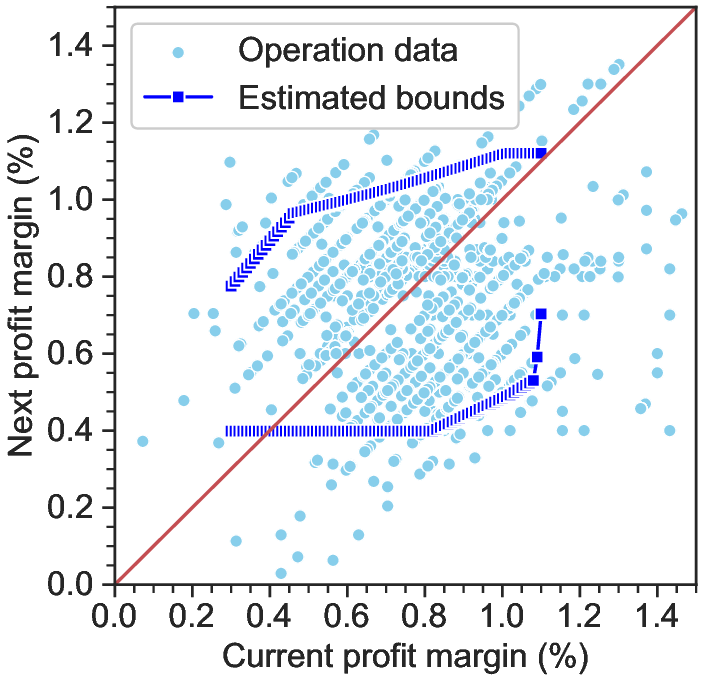}
    \subcaption{Product C}
  \end{minipage}
  \caption{Estimated bounds of the profit margin for each product using the DM-based approach with the minimum support of 1.0 and step size of $0.01\%$} \label{fig:abc}
\end{figure}

Fig.~\ref{fig:abc} presents the results of adjusting the profit margin bounds for each product derived using the DM-based approach by incorporating both monotonicity and convexity-concavity constraints.
The proposed adjustment approach facilitates the following interpretation of the trends in the profit margin bounds.
\begin{itemize}
\item \textbf{Product A:} When the current profit margin exceeds approximately 0.6\%, the upper bounds tend to be constrained. Conversely, when it falls below this threshold, the lower bounds are similarly restrained. Despite this behavior, the upper and lower bounds remain roughly symmetrical. This pattern aligns with the characteristics of the product, where the influence of competing products is limited, and pricing decisions are made in response to business conditions.
\vspace{2mm}
\item \textbf{Product B:} Both the lower and upper bounds of the profit margin are tight. Specifically, the upper bounds tend to be restrictive, resulting in infrequent price increases. This is consistent with the characteristics of the product, where price increases are challenging due to the significant impact of competing products.
\vspace{2mm}
\item \textbf{Product C:} 
The lower bounds of the profit margin remain independent of the current profit margin up to approximately 0.8\%. Beyond this point, the lower bounds increase proportionally with the profit margin, though the slope is gradual. This behavior reflects the characteristics of the product, which is less seasonal than others and can only be reduced to a certain extent due to its constant demand.
\end{itemize}

\section{Conclusion}
We proposed an interpretable framework for determining price bounds consisting of two main components: price-bound estimation and adjustment, using historical pricing data.
First, we introduced three approaches for estimating the profit margin bounds: NR-, DM-, and ML-based approaches.
Subsequently, we presented a method for adjusting the estimated bounds through mathematical optimization.

In numerical experiments with real pricing data from Recruit Co., Ltd., we found that the proposed method improved estimation accuracy and that the adjusted bounds increased stability, making them less susceptible to outliers and easier for operators to interpret.
These improvements are expected to enhance the validity of pricing strategies and facilitate strategic decision-making by providing clearer insights into pricing practices.
Additionally, our framework can be readily applied to personalized pricing \cite{biggs2021model,ikeda2024robust}.

A segment of our framework—specifically the DM-based estimation approach combined with adjustments via mathematical optimization—was implemented in the price-optimization system at Recruit Co., Ltd.
This implementation greatly facilitated the seamless integration of the system into actual pricing operations and improved the interpretation of price bounds.
As a result, this led to considerable savings in labor costs in pricing operations and an increase in revenue and profits for the services.


%
\section*{Declarations}
\subsection*{Conflict of interest}
On behalf of all the authors, the corresponding author states that there are no conflicts of interest.

\bibliographystyle{spmpsci}    
\bibliography{ref}   


\end{document}